\documentclass[aps,pra,twocolumn,groupedaddress,showpacs]{revtex4-1}
\usepackage{graphicx}
\usepackage{color}
\usepackage{amsmath}
\usepackage{lipsum}

\usepackage[normalem]{ulem}

\begin{document}


\title{Hidden momentum of electrons, nuclei, atoms and molecules}

\author{Robert P. Cameron}
\email{robert.p.cameron@strath.ac.uk, www.ytilarihc.com}
\address{SUPA and Department of Physics, University of Strathclyde, Glasgow G4 0NG, U.K.}

\author{J. P. Cotter}
\address{Centre for Cold Matter, Blackett Laboratory, Imperial College London, Prince Consort Road, London SW7 2AZ, U.K.}

\begin{abstract}
We consider the positions and velocities of electrons and spinning nuclei and demonstrate that these particles harbour hidden momentum when located in an electromagnetic field. This hidden momentum is present in all atoms and molecules, however it is ultimately cancelled by the momentum of the electromagnetic field. We point out that an electron vortex in an electric field might harbour a comparatively large hidden momentum and recognise the phenomenon of hidden \textit{hidden} momentum.
\end{abstract}

\date{\today}
\maketitle


\section{Introduction}
\begin{figure}[h!]
\centering
\includegraphics[width=0.8\linewidth]{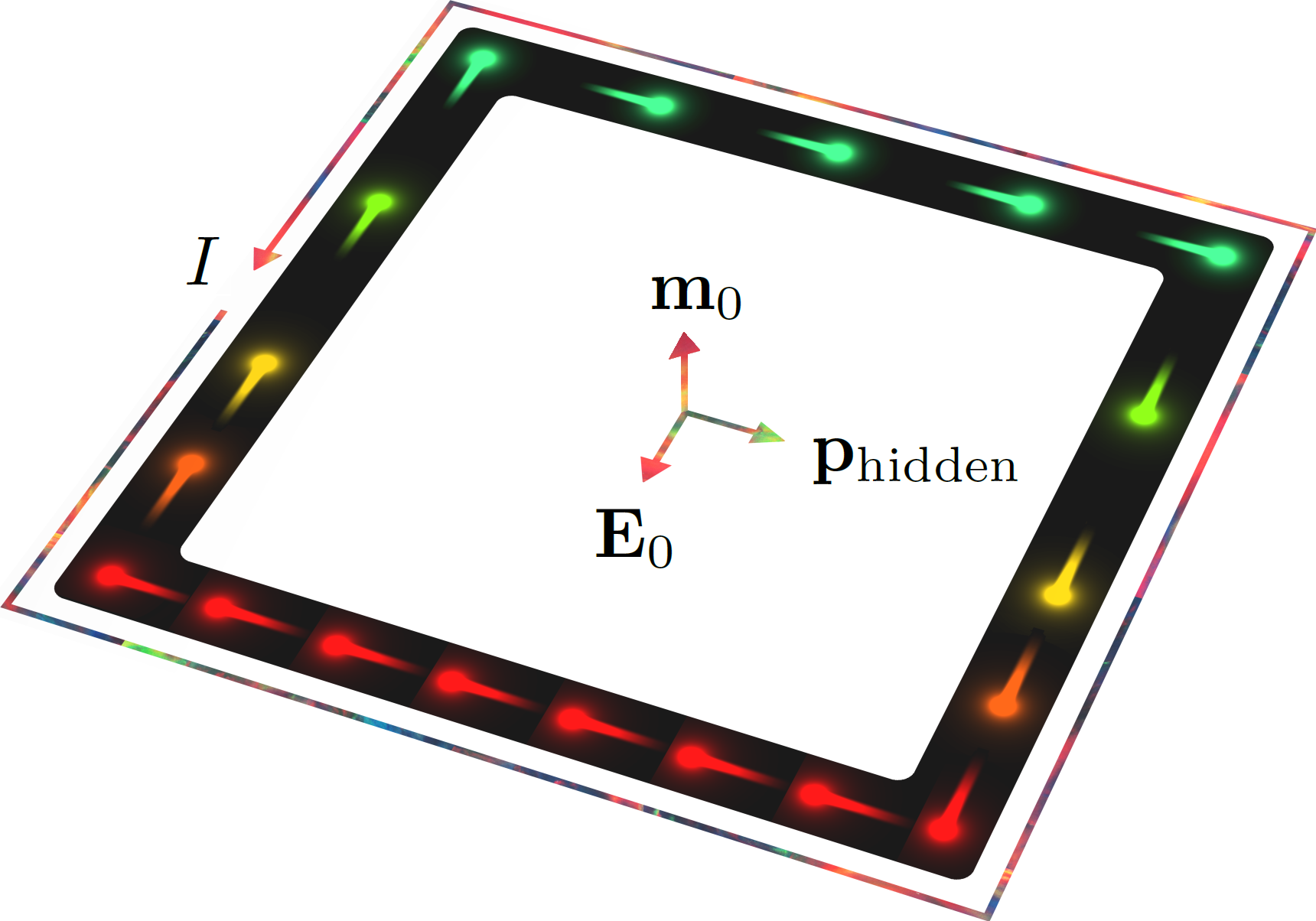}\\
\caption{\small A loop of electric current in a static electric field. Different charge carriers in the loop have different speeds, as indicated here by their colour. The imbalance of their momenta is the loop's hidden momentum\,\cite{Shockley67a,Griffiths99, Babson09a, Filho15a}.}
\label{Hiddenmomentum1}
\end{figure}

A loop of electric current $I$ and magnetic-dipole moment $\mathbf{m}_0$ at rest in a static electric field $\mathbf{E}_0$ has `hidden momentum' $\mathbf{p}_\textrm{hidden}=\mathbf{m}_0\times\mathbf{E}_0/c^2$, even though the loop is not moving\,\cite{Shockley67a, Griffiths99, Babson09a, Filho15a}. This system is illustrated in FIG.\,\ref{Hiddenmomentum1}. The hidden momentum results from the different charge carriers in the loop having different speeds, due to a modification of their usual motion around the loop by $\mathbf{E}_0$\,\cite{Griffiths99, Babson09a, Filho15a}. It is cancelled by the momentum $-\mathbf{m}_0\times\mathbf{E}_0/c^2$ of the electromagnetic field\,\cite{Thomson04a,Thomson04b,Thomson04c, Griffiths99, Babson09a}. The phenomenon of hidden momentum is not unique to this system, nor is it unique to electrodynamics\,\cite{Shockley67a, vanVleck69a, Griffiths99, Babson09a}.

This paper was motivated by a question posed recently by Filho and Saldanha: ``does an electron with a magnetic moment resulting from its spin in the presence of an applied electric field have hidden momentum"\,\cite{Filho15a}? In \S\ref{Free electron}, we consider a free electron, as described by the (first-quantised) Dirac equation. We highlight subtleties associated with `the' position and velocity of the electron, an understanding of which is necessary for the analysis that follows. In \S\ref{Electron in an external electromagnetic field}, we introduce an external electromagnetic field and demonstrate that the electron harbours hidden momentum associated with its spin, thus providing an affirmative answer to the question above. In \S\ref{Isolated atom or molecule}, we consider an isolated atom or molecule and reaffirm that its constituent electrons, as well as any spinning nuclei present, harbour hidden momentum individually. We also show that the sum total of this hidden momentum is cancelled by the momentum of the electromagnetic field, as it should be. In \S\ref{Outlook}, we point out that an electron vortex in an electric field might harbour a comparatively large hidden momentum and recognise the hitherto neglected phenomenon of hidden \textit{hidden} momentum. Our work is timely, given the recent surge of interest in relativistic electron vortices\,\cite{Bliokh11a,Larocque16a,Bialynicki-Birula17a,Barnett17a,Bliokh17aa,Kruining17a,Lloyd17a,Bliokh17a}.  

In what follows, `hats' are used to indicate physical quantities whereas the mathematical operators used to express these quantities in different representations do not have hats -- we alternate between the Dirac representation (primed)\,\cite{Dirac28a} and the Foldy-Wouthuysen representation (unprimed)\,\cite{Foldy50a}, defined in appendix \ref{Representations}. These distinctions are important. Consider, for example, $\hat{\mathbf{r}}_q'=\hat{\mathbf{r}}_M=\mathbf{r}$. Here, two different physical quantities ($\hat{\mathbf{r}}_q'$ and $\hat{\mathbf{r}}_M$) are expressed in two different representations (primed and unprimed) by the same mathematical operator ($\mathbf{r}$).


\section{Free electron}
\label{Free electron}
Let us consider first a \textit{free electron}. In the Dirac representation, the electron obeys
\begin{equation}
\textrm{i}\hbar\dot{\psi}'=\hat{H}'\psi',
\end{equation}
with $\psi'=\psi'(\mathbf{r},t)$ the electron's spinor and
\begin{equation}
\hat{H}'=c \pmb{\alpha}\cdot \mathbf{p}+\beta m c^2
\end{equation}
the free Dirac Hamiltonian\,\cite{Dirac28a}. Here,
\begin{eqnarray}
\pmb{\alpha}=\left(\begin{array}{c c}
0 & \pmb{\sigma} \\
\pmb{\sigma} & 0
\end{array}\right), \ \   \mathbf{p}=-\textrm{i}\hbar\pmb{\nabla},  \ \ \beta=\left(\begin{array}{c c}
1 & 0 \\
0 & -1
\end{array}\right),
\end{eqnarray} 
and $m$ is the rest mass of the electron. The momentum of the electron can be identified unambiguously as $\hat{\mathbf{p}}'=\hat{\mathbf{p}}=\mathbf{p}$. However, `the' position and velocity of the electron are not unique\,\cite{Born35a,Pryce35a,Pryce48a,Foldy50a,Barut68a,vanVleck69a,Barone73a,Barut81a,Bliokh11a,Bliokh12a,Bliokh17a,Smirnova18a}. For the purposes of this paper, we find it \textit{necessary} to identify and distinguish between the instantaneous position $\hat{\mathbf{r}}'_q$ of the electron's electric charge, the kinetic position $\hat{\bar{\mathbf{r}}}_q'$ of the electron and the so-called mean position $\hat{\mathbf{r}}'_M$ of the electron. As the electron is free, these positions can be defined as follows.

\subsection{Positions and velocities}
The position of charge takes on a simple form in the Dirac representation\,\cite{Dirac28a},
\begin{eqnarray}
\hat{\mathbf{r}}'_q=\mathbf{r}.
\end{eqnarray}
The interpretation of $\hat{\mathbf{r}}'_q$ as the position of charge\,\cite{Barut68a,Barut81a} will be made apparent in the next section, where we impose an electromagnetic field.

The kinetic position is\,\cite{Schrodinger30a,Schrodinger31a,Born35a}
\begin{eqnarray}
\hat{\bar{\mathbf{r}}}'_q&=&\frac{1}{4}\left[\frac{1}{\hat{H}'}\left(\hat{H}'\hat{\mathbf{r}}_q'+\hat{\mathbf{r}}_q'\hat{H}'\right)+\left(\hat{H}'\hat{\mathbf{r}}_q'+\hat{\mathbf{r}}_q'\hat{H}'\right)\frac{1}{\hat{H}'}\right]. \nonumber \\
&~&
\end{eqnarray}
This coincides with the centre of the electron's electric charge, as evidenced by the result that $\langle\hat{\bar{\mathbf{r}}}'_q\rangle=\langle\hat{\mathbf{r}}_q'\rangle$ for a state with energy of definite sign\,\cite{Bliokh17a}. $\hat{\bar{\mathbf{r}}}'_q$ is sometimes referred to as the `observable' part of the position $\hat{\mathbf{r}}'_q$ of charge\,\cite{Pryce48a}, being the projection of $\hat{\mathbf{r}}'_q$ onto positive and negative energy subspaces\,\cite{Bliokh11a,Bliokh17a}. $\hat{\bar{\mathbf{r}}}'_q$ is not the electron's centre of energy \cite{Bliokh17a}, in spite of its suggestive form.

The mean position takes on a simple form in the Foldy-Wouthuysen representation\,\cite{Pryce35a,Foldy50a},
\begin{equation}
\hat{\mathbf{r}}_M=\mathbf{r}.
\end{equation}
Loosely speaking, $\hat{\mathbf{r}}_M$ can be thought of as the kinetic position in the electron's rest frame, actively boosted with appropriate velocity\,\cite{Pryce48a,Giambiagi60a,Saavedra65a,Bliokh17a}. It is $\hat{\mathbf{r}}_M$ that is usually regarded as being `the' position of the electron in low-energy studies\,\cite{Pauli27a,Foldy50a}, although one can argue that the kinetic position $\hat{\bar{\mathbf{r}}}_q$ is closer to the classical notion of position for a particle like the electron\,\cite{Barone73a}. The `mean' terminology introduced in\,\cite{Foldy50a} for $\hat{\mathbf{r}}_M$ and other quantities is something of a misnomer -- it is $\hat{\bar{\mathbf{r}}}_q$ rather than $\hat{\mathbf{r}}_M$ that embodies the electron's `average' position\,\cite{Barut81a}.

The components of the velocity $\hat{\mathbf{v}}'_q=\textrm{d}\hat{\mathbf{r}}'_q/\textrm{d}t=c\pmb{\alpha}$ of charge\,\cite{Eddington29a,Breit29a} support discrete eigenvalues of $\pm c$ whilst the kinetic velocity $\hat{\bar{\mathbf{v}}}_q=\textrm{d}\hat{\bar{\mathbf{r}}}_q/\textrm{d}t=\beta c^2 \mathbf{p}/ E_p$\,\cite{Barone73a} and mean velocity $\hat{\mathbf{v}}_M=\textrm{d}\hat{\mathbf{r}}_M/\textrm{d}t=\beta c^2 \mathbf{p}/ E_p$\,\cite{Foldy50a} vary continuously with $\mathbf{p}$ and are equal. Here,
\begin{equation}
E_p=\sqrt{m^2c^4+c^2 p^2}.
\end{equation}

The above can be summarised as follows:
\begin{equation}
\begin{tabular}{ c | c } 
\textrm{quantity} & \textrm{definition} \\ \hline
\textrm{position of charge} & $\hat{\mathbf{r}}_q'=\mathbf{r}$   \\ 
\textrm{kinetic position} & $\hat{\bar{\mathbf{r}}}_q'=\frac{1}{4}\left\{\frac{1}{\hat{H}'},\left\{\hat{H}',\hat{\mathbf{r}}_q'\right\}\right\}$  \\
\textrm{mean position} & $\hat{\mathbf{r}}_M=\mathbf{r}$ \\ 
\textrm{velocity of charge} & $\hat{\mathbf{v}}_q'=c\pmb{\alpha}$  \\ 
\textrm{kinetic velocity} & $\hat{\bar{\mathbf{v}}}_q=\beta c^2\mathbf{p}/E_p$ \\ 
\textrm{mean velocity} & $\hat{\mathbf{v}}_M=\beta c^2\mathbf{p}/E_p$ \\ 
\textrm{momentum} & $\hat{\mathbf{p}}'=\hat{\mathbf{p}}=\mathbf{p}$
\end{tabular}
\end{equation}
where we have used curly brackets to indicate anti-commutators.

\subsection{\textit{Zitterbewegung}}
In the Heisenberg picture, the positions evolve as\,\cite{Schrodinger30a,Schrodinger31a,Barut81a}
\begin{eqnarray}
\hat{\mathbf{r}}_q(t)&=&\hat{\bar{\mathbf{r}}}_q(t)+\hat{\pmb{\xi}}(t), \\
\hat{\bar{\mathbf{r}}}_q(t)&=&\hat{\mathbf{r}}_M(t)+\hat{\pmb{\delta}}, \\
\hat{\mathbf{r}}_M(t)&=&\hat{\mathbf{r}}_M(0)+\hat{\mathbf{v}}_M t,
\end{eqnarray}
with
\begin{eqnarray}
\hat{\pmb{\xi}}(t)&=&\frac{\textrm{i}\hbar c}{2}\left[\hat{\mathbf{v}}_q(t)-\frac{c\hat{\mathbf{p}}}{\hat{H}}\right]\frac{\textrm{e}^{-2\textrm{i}\hat{H}t/\hbar}}{\hat{H}}, \\
\hat{\pmb{\delta}}&=&\frac{c^2 \mathbf{p}\times\mathbf{s}}{E_p (E_p+mc^2)} \nonumber \\
&=&\frac{\mathbf{p}\times\mathbf{s}}{2 m^2 c^2}+\textrm{O}(\tfrac{1}{c^3}). \label{deltafree} 
\end{eqnarray}
Here,
\begin{equation}
\mathbf{s}=\frac{\hbar}{2}\left(\begin{array}{c c}
\pmb{\sigma}  & 0 \\
0 & \pmb{\sigma}
\end{array}\right).
\end{equation}
The position difference $\hat{\pmb{\xi}}(t)$ executes a complicated oscillatory motion with amplitude comparable to the Compton wavelength $2\pi \hbar /mc$ and the resulting motion of the position $\hat{\mathbf{r}}_q(t)$ of charge is referred to as the electron's \textit{Zitterbewegung}\,\cite{Schrodinger30a, Schrodinger31a, Foldy50a, Huang52a, Barut81a}. Meanwhile, the kinetic position $\hat{\bar{\mathbf{r}}}_q(t)$ and the mean position $\hat{\mathbf{r}}_M(t)$ translate uniformly, with $\hat{\bar{\mathbf{r}}}_q(t)$ offset from $\hat{\mathbf{r}}_M(t)$ by the position difference $\hat{\pmb{\delta}}$. The equality $\hat{\bar{\mathbf{v}}}_q=\textrm{d}\hat{\bar{\mathbf{r}}}_q(t)/\textrm{d}t=\hat{\mathbf{v}}_M=\textrm{d}\hat{\mathbf{r}}_M(t)/\textrm{d}t=\beta c^2 \mathbf{p}/ E_p$ holds as $\hat{\pmb{\delta}}$ is constant.

\subsection{Relativistic Hall effect}
\begin{figure}[h!]
\centering
\includegraphics[width=\linewidth]{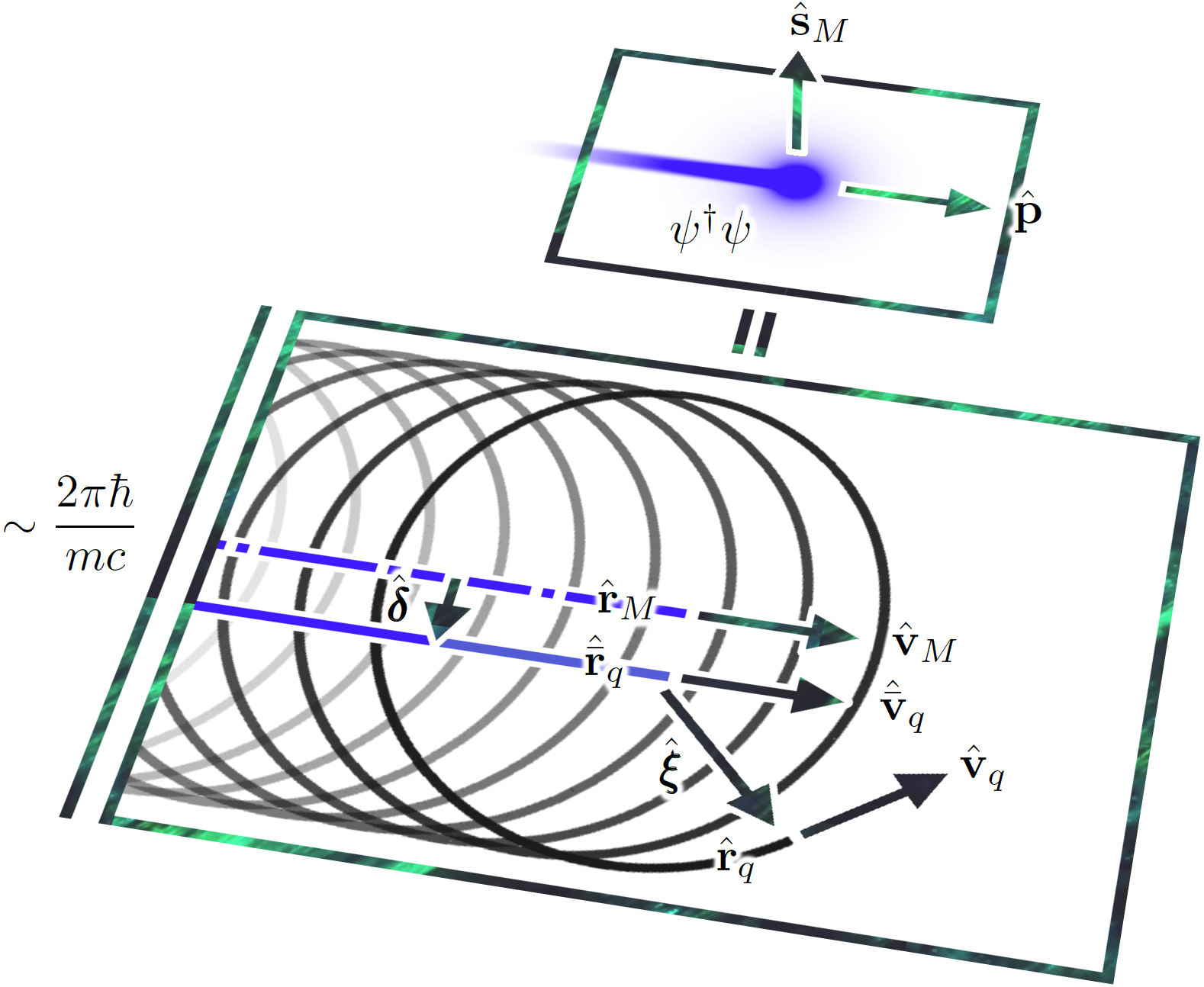}\\
\caption{\small 
There is a sense in which a free electron resembles an electric current `loop' -- the electron's \textit{Zitterbewegung} sees the position of charge circulate, in spite of there being no obvious external fields. It can be argued that this is the origin of the electron's spin and magnetic-dipole moment\,\cite{Schrodinger30a, Schrodinger31a, Foldy50a, Huang52a, Barut81a}. It seems natural, therefore, to anticipate that an electron in an electric field harbours hidden momentum associated with its spin, due to a modification of its \textit{Zitterbewegung} by the field.}
\label{Hiddenmomentum2}
\end{figure}

The position difference $\hat{\pmb{\delta}}$ might be regarded as a manifestation of the relativistic Hall effect\,\cite{Bliokh12a}. It embodies a distortion of the trajectory of the electron's charge, due to the electron's spin and translation. A rotating, translating wheel serves as an instructive (classical) analogy -- different elements on the rim ($\hat{\mathbf{r}}_q$) have different speeds and are Lorentz contracted by different amounts, giving a shift ($\hat{\pmb{\delta}}$) of the element-weighted centre ($\hat{\bar{\mathbf{r}}}_q$) away from the axle ($\hat{\mathbf{r}}_M$)\,\cite{Muller92a,Bliokh12a}. Due to the spin dependence of $\hat{\pmb{\delta}}$, the components of $\hat{\bar{\mathbf{r}}}_q$ do not commute: $[\hat{\bar{r}}_{q\alpha},\hat{\bar{r}}_{q\beta}]=-\textrm{i}\hbar c^2 \epsilon_{\alpha\beta\gamma}\hat{\bar{s}}_\gamma/\hat{H}^{2}$\,\cite{Born35a}. Here, $\hat{\bar{\mathbf{s}}}=\hat{\mathbf{j}}-\hat{\bar{\mathbf{r}}}_q\times\hat{\mathbf{p}}$\,\cite{Schrodinger30a, Schrodinger31a}, with $\hat{\mathbf{j}}$ the total angular momentum of the electron\,\cite{Dirac28a}. It seems that $\hat{\bar{\mathbf{s}}}$ is the electronic analogue\,\cite{Bliokh11a,Bliokh17a} of the spin of freely propagating light\,\cite{Darwin32a,vanEnk94a,vanEnk94b,Barnett10a,Bliokh10a,Ivo11a} -- each is conserved and both have similar commutations relations. $\hat{\mathbf{r}}_q(t)$, $\hat{\bar{\mathbf{r}}}_q(t)$ and $\hat{\mathbf{r}}_M(t)$ are depicted \textit{schematically} in FIG.\,\ref{Hiddenmomentum2}. 

Position differences like $\hat{\pmb{\delta}}$ are well known for electrons in the solid state and can be regarded as Berry connections in momentum space \cite{Adams59a,Bliokh05a,Berard06a,Chang08a,Xiao10a,Bliokh11a,Takahashi15a, Bliokh17a}.

\section{Electron in an external electromagnetic field}
\label{Electron in an external electromagnetic field}
To demonstrate that an electron can harbour hidden momentum associated with its spin, let us consider now an \textit{electron in an external electromagnetic field}, with scalar potential $\Phi=\Phi(\mathbf{r},t)$ and magnetic vector potential $\mathbf{A}=\mathbf{A}(\mathbf{r},t)$ in the Coulomb gauge\,\cite{Maxwell61a}. We work to order $1/c^2$ and assume that the leading-order contribution to $\mathbf{A}$ is of order $1/c^2$. According to the principle of minimal coupling, the Hamiltonian in the Dirac representation becomes \cite{Dirac28a}
\begin{equation}
\hat{H}'=c \pmb{\alpha}\cdot (\mathbf{p}-q\mathbf{A})+\beta m c^2+q\Phi \label{Diracminimal},
\end{equation}
where $q$ is the electron's electric charge. It is important now to distinguish between the canonical momentum $\hat{\mathbf{p}}'=\mathbf{p}=\hat{\mathbf{p}}+\textrm{O}(1/c^3)$ and the total kinetic momentum $\hat{\pmb{\pi}}'=\mathbf{p}-q\mathbf{A}=\hat{\pmb{\pi}}+\textrm{O}(1/c^3)$ of the electron. It is $\hat{\pmb{\pi}}'$ rather than $\hat{\mathbf{p}}'$ that obeys the Lorentz force law\,\cite{Lorentz95a,Eddington29a,Breit29a},
\begin{equation}
\frac{\textrm{d}\hat{\pmb{\pi}}'}{\textrm{d}t}=q(\mathbf{E}+\pmb{\alpha}\times\mathbf{B}). \label{forcelaw}
\end{equation}
Here, $\mathbf{E}=\mathbf{E}(\mathbf{r},t)=-\pmb{\nabla}\Phi-\dot{\mathbf{A}}$ is the electric field and $\mathbf{B}=\mathbf{B}(\mathbf{r},t)=\pmb{\nabla}\times\mathbf{A}$ is the magnetic field. The absence of explicit magnetic-dipole moment terms in (\ref{forcelaw}) agrees with the view that the magnetic-dipole moment of the electron is an emergent feature, due to the electron's \textit{Zitterbewegung}\,\cite{Schrodinger30a, Schrodinger31a, Foldy50a, Huang52a, Barut81a}.

\subsection{Positions and velocities}
We continue to identify and distinguish between the position $\hat{\mathbf{r}}'_q$ of charge, the kinetic position $\hat{\bar{\mathbf{r}}}_q$ and the mean position $\hat{\mathbf{r}}_M$ -- as the electron is in the presence of an electromagnetic field, we now define these as
\begin{eqnarray}
\hat{\mathbf{r}}'_q=\mathbf{r}, \ \  \hat{\bar{\mathbf{r}}}_q=\hat{\mathbf{r}}_M+\hat{\pmb{\delta}}, \ \ \& \ \ \hat{\mathbf{r}}_M=\mathbf{r},
\end{eqnarray}
with
\begin{equation}
\hat{\pmb{\delta}}=\frac{\mathbf{p}\times\mathbf{s}}{2m^2c^2}+\textrm{O}(\tfrac{1}{c^3}).
\end{equation}
Note that the potentials in (\ref{Diracminimal}) are evaluated at $\mathbf{r}$, in accord with the interpretation of $\hat{\mathbf{r}}_q'$ as the position of charge\,\cite{Barut68a,Barut81a}. Unlike the case for a free electron, the position difference $\hat{\pmb{\delta}}$ is not necessarily constant -- the electromagnetic field can alter $\mathbf{p}\times\mathbf{s}$ to leading order, as $\textrm{d}\hat{\mathbf{p}}/\textrm{d}t=-q\pmb{\nabla}\Phi+\textrm{O}(1/c^1)$ and $\textrm{d}\hat{\mathbf{s}}_M/\textrm{d}t=0+\textrm{O}(1/c^1)$, with $\hat{\mathbf{s}}_M=\mathbf{s}$ the mean spin of the electron\,\cite{Foldy50a}. It follows that the kinetic velocity $\hat{\bar{\mathbf{v}}}_q=\textrm{d}\hat{\bar{\mathbf{r}}}_q/\textrm{d}t$ no longer equals the mean velocity $\hat{\mathbf{v}}_M=\textrm{d}\hat{\mathbf{r}_M}/\textrm{d}t$, as
\begin{equation}
\hat{\bar{\mathbf{v}}}_q-\hat{\mathbf{v}}_M=\frac{\textrm{d}\hat{\pmb{\delta}}}{\textrm{d}t}=\frac{q \mathbf{s}\times\pmb{\nabla}\Phi}{2m^2c^2}. \label{velocitydifferenceinfield}
\end{equation}
This subtlety will prove important below.

Velocity contributions like $\textrm{d}\hat{\pmb{\delta}}/\textrm{d}t$ are also known for electrons in the solid state and are sometimes referred to as being `anomalous' \cite{Adams59a,Bliokh05a,Berard06a,Chang08a,Xiao10a,Takahashi15a}.

\subsection{Hidden momentum}
Explicit calculation of $\textrm{i}[\hat{H},\hat{\mathbf{r}}_M]/\hbar$ reveals that the mean velocity is
\begin{equation}
\hat{\mathbf{v}}_M=\frac{\beta\mathbf{p}}{m}-\frac{\beta p^2 \mathbf{p}}{2m^3c^2}-\beta q \mathbf{A}+\frac{q \mathbf{s}\times\pmb{\nabla}\Phi}{2m^2 c^2}+\textrm{O}(\tfrac{1}{c^3}). \label{step1}
\end{equation}
Multiplying this by $\beta$ and rearranging reveals that the canonical momentum is
\begin{eqnarray}
\hat{\mathbf{p}}&=&\beta m \hat{\mathbf{v}}_M+\frac{\hat{p}^2\hat{\mathbf{p}}}{2m^2c^2}+q\mathbf{A}-\frac{\beta q(\mathbf{s}\times\pmb{\nabla}\Phi )}{2mc^2}+\textrm{O}(\tfrac{1}{c^3}).  \nonumber \\
&~&\label{step2}
\end{eqnarray}
It is tempting to identify the first and second terms here with the relativistically corrected kinetic momentum, and the third term with the electromagnetic momentum. However, this leaves the fourth term unaccounted for. To proceed, we must recognise that the kinetic momentum should be cast in terms of the kinetic velocity $\hat{\bar{\mathbf{v}}}_q$, rather than the mean velocity $\hat{\mathbf{v}}_M$. This leads us to recast the spin-dependent term in (\ref{step2}) as
\begin{eqnarray}
&-&\frac{\beta q(\mathbf{s}\times\pmb{\nabla}\Phi)}{2mc^2}=\frac{\beta q(\mathbf{s}\times\pmb{\nabla}\Phi)}{2mc^2} -\frac{\beta q(\mathbf{s}\times\pmb{\nabla}\Phi)}{mc^2}\nonumber \\
&=&\beta m (\hat{\bar{\mathbf{v}}}_q-\hat{\mathbf{v}}_M)+\frac{\hat{\mathbf{m}}\times\hat{\mathbf{E}}}{c^2}+\textrm{O}(\tfrac{1}{c^3}), \label{step3}
\end{eqnarray}
with $\hat{\mathbf{m}}=\beta q \mathbf{s} /m$ the magnetic-dipole moment of the electron. Here, we have made use of (\ref{velocitydifferenceinfield}) and $\mathbf{E}=-\pmb{\nabla}\Phi+\textrm{O}(1/c^1)$. Substituting (\ref{step3}) into (\ref{step2}) gives
\begin{eqnarray}
\hat{\mathbf{p}}&=&\beta m\hat{\bar{\mathbf{v}}}_q+\frac{\hat{p}^2\hat{\mathbf{p}}}{2m^2c^2}+q \mathbf{A}+\frac{\hat{\mathbf{m}}\times\mathbf{E}}{c^2}+\textrm{O}(\tfrac{1}{c^3}). \label{finalpsingle}
\end{eqnarray}
Thus, $\hat{\mathbf{p}}$ is comprised of relativistically corrected kinetic momentum terms ($\beta m \hat{\bar{\mathbf{v}}}_q+\hat{p}^2\hat{\mathbf{p}}/2m^2c^2$), an electromagnetic momentum term ($q\mathbf{A}$) and, pleasingly, a hidden momentum term ($\hat{\mathbf{m}}\times\mathbf{E}/c^2$) with the prototypical form described in the introduction. We attribute this hidden momentum to a modification of the electron's \textit{Zitterbewegung} by the electric field $\mathbf{E}$. For the special case in which $\mathbf{E}$ is due to a ``test particle'', a complementary result was derived in\,\cite{vanVleck69a}. A similar result was derived in\,\cite{Barone73a}, but with no explicit recognition of the hidden momentum. Note that the total kinetic momentum $\hat{\pmb{\pi}}$ includes the hidden momentum $\hat{\mathbf{m}}\times\mathbf{E}/c^2$.

The hidden momentum $\hat{\mathbf{m}}\times\mathbf{E}/c^2$ is small, its expectation value being $\lesssim-q\hbar |\mathbf{E}|/2mc^2=|\mathbf{E}|\times 10^{-40}\textrm{A}.\textrm{s}^2$ in magnitude.


\section{Isolated atom or molecule}
\label{Isolated atom or molecule}
The formalism employed in the previous section does not allow us to confirm that the hidden momentum $\hat{\mathbf{m}}\times\mathbf{E}/c^2$ is cancelled by the momentum of the electromagnetic field, as the field is externally imposed. Let us conclude, therefore, by considering \textit{an isolated atom or molecule} -- a closed system. Our description is effectively truncated at order $1/c^2$ and we therefore neglect terms of order $1/c^3$ or smaller. The subscripted `$q$' and `$M$' notation used above is henceforth dropped, for the sake of clarity. Let us focus our discussion upon a molecule (an atom being a special case with one nucleus). We regard the molecule as being an electrically neutral collection of electrons (subscript $i$) and spin $0$ or $1/2$ nuclei (subscript $j$), bound together by electromagnetic interactions in the absence of external influences. We refer to the electrons and nuclei collectively as `the particles' (subscript $k$) and treat the $k$th particle as a point-like object of rest mass $m_k$, mean position $\hat{\mathbf{r}}_k=\mathbf{r}_k$, canonical momentum $\hat{\mathbf{p}}_k=-\textrm{i}\hbar\pmb{\nabla}_k$, electric charge $q_k$ and magnetic-dipole moment $\hat{\mathbf{m}}_k=\gamma_k \hat{\mathbf{s}}_k$, with $\gamma_k$ the gyromagnetic ratio and $\hat{\mathbf{s}}_k=\hbar\pmb{\sigma}_k/2$ the mean spin, where it is to be understood that $\hat{\mathbf{m}}_k/\gamma_k=\hat{\mathbf{s}}_k=0$ for spin $0$ nuclei. Let $R_i=\sqrt{3}\hbar/2m_ic$ account for the effective finite sizes of the electrons\,\cite{Darwin28a,Itoh65a}, $R_j$ account for the finite size of the $j$th nucleus\,\cite{Pachucki95a,Helgaker} and $f_k=\left(1-q_k/2m_k\gamma_k\right)$ be the usual spin-orbit factor\,\cite{Uhlenbeck26a, Thomas26a, Thomas27a, Gunther-Mohr54a} for the $k$th particle. We regard the $R_k^2$ as being of order $1/c^2$ and take the Hamiltonian governing our molecule to be\,\cite{Breit29a,Breit30a, Breit32a,Chraplyvy53a, Chraplyvy53b,Itoh65a,Pachucki95a,Helgaker}
\begin{eqnarray}
\hat{H}&=&\sum_k\frac{\hat{p}_k^2}{2m_k}+\sum_k \frac{1}{2}q_k \hat{\Phi}_k-\sum_k\frac{\hat{p}_k^4}{8m_k^3c^2} \nonumber  \\
&+&\sum_k \frac{1}{6} q_k R_k^2 \nabla^2_k \hat{\Phi}^q_k-\sum_k\frac{f_k\hat{\mathbf{m}}_k\cdot(\hat{\mathbf{p}}_k\times\pmb{\nabla}_k\hat{\Phi}_k^q)}{m_kc^2} \nonumber \\
&-&\sum_k\frac{q_k\hat{\mathbf{p}}_k\cdot \hat{\mathbf{A}}_k}{2m_k}-\sum_k\frac{1}{2} \hat{\mathbf{m}}_k\cdot(\pmb{\nabla}_k\times\hat{\mathbf{A}}_k),
\end{eqnarray}
with
\begin{equation}
\hat{\Phi}_k=\hat{\Phi}_k^q+\hat{\Phi}_k^R
\end{equation}
the intramolecular scalar potential seen by the $k$th particle at $\hat{\mathbf{r}}_k$ and
\begin{equation}
\hat{\mathbf{A}}_k=\hat{\mathbf{A}}_k^{\mathbf{m}}+\hat{\mathbf{A}}_k^{\mathbf{v}}
\end{equation}
the intramolecular magnetic vector potential, where
\begin{eqnarray}
\hat{\Phi}_k^q&=&\sum_{k'\ne k}\frac{q_{k'}}{4\pi\epsilon_0\hat{r}_{kk'}} \\
\hat{\Phi}_k^R&=&-\sum_{k'\ne k}\frac{q_{k'}R_{k'}^2\delta^3\left(\hat{\mathbf{r}}_{kk'}\right)}{6\epsilon_0}
\end{eqnarray}
account for the electric charges and finite sizes of the other particles and
\begin{eqnarray}
\hat{\mathbf{A}}_k^{\mathbf{m}}&=&\sum_{k'\ne k}\frac{\mu_0\hat{\mathbf{m}}_{k'}\times\hat{\mathbf{r}}_{kk'}}{4\pi\hat{r}_{kk'}^3} \\
\hat{\mathbf{A}}_k^{\mathbf{v}}&=&\sum_{k'\ne k}\frac{\mu_0 q_{k'}}{16\pi m_{k'}}\Bigg[\frac{1}{\hat{r}_{kk'}}\hat{\mathbf{p}}_{k'}+\hat{\mathbf{p}}_{k'}\frac{1}{\hat{r}_{kk'}} \nonumber \\
&+&\hat{\mathbf{r}}_{kk'}\frac{1}{\hat{r}_{kk'}^3}\left(\hat{\mathbf{r}}_{kk'}\cdot\hat{\mathbf{p}}_{k'}\right)+(\hat{\mathbf{p}}_{k'}\cdot\hat{\mathbf{r}}_{kk'})\frac{1}{\hat{r}_{kk'}^3}\hat{\mathbf{r}}_{kk'}\Bigg]
\end{eqnarray}
account for the intrinsic magnetic moments and orbital motions.

\subsection{Hidden momentum of the electrons and nuclei individually}
Defining\,\footnote{For the $j$th nucleus, the kinetic position $\hat{\bar{\mathbf{r}}}_j$ appears to differ from the centre $\hat{\mathbf{r}}_j+(2m_j\gamma_j/q_j-1)\hat{\pmb{\delta}}_j$ of charge.}
\begin{eqnarray}
\hat{\pmb{\delta}}_k&=&\hat{\bar{\mathbf{r}}}_k-\hat{\mathbf{r}}_k=\frac{\hat{\mathbf{p}}_k\times\hat{\mathbf{s}}_k}{2 m_k^2 c^2}, \\
\hat{\bar{\mathbf{v}}}_k&=&\frac{\textrm{d}\hat{\bar{\mathbf{r}}}_k}{\textrm{d}t}, \\
\hat{\mathbf{v}}_k&=&\frac{\textrm{d}\hat{\mathbf{r}}_k}{\textrm{d}t},
\end{eqnarray}
a calculation analogous to that outlined in the previous section reveals that the canonical momentum of the $k$th particle is
\begin{eqnarray}
\hat{\mathbf{p}}_k&=&m_k\hat{\bar{\mathbf{v}}}_k+\frac{\hat{p}_k^2\hat{\mathbf{p}}_k}{2m_k^2c^2}+q_k \hat{\mathbf{A}}_k-\frac{\hat{\mathbf{m}}_k\times\pmb{\nabla}_k\hat{\Phi}_k^q}{c^2}+\textrm{O}(\tfrac{1}{c^3}). \nonumber \\
\label{finalpmulti} 
\end{eqnarray}
Thus, each electron and spinning nucleus in the molecule harbours a hidden momentum $-\hat{\mathbf{m}}_k\times\pmb{\nabla}_k\hat{\Phi}_k^q/c^2$. 

A basic estimate suggests that the hidden momentum of an electron in a hydrogen atom corresponds to a notional electronic speed of only $\lesssim 5\times 10^1\textrm{m}.\textrm{s}^{-1}$. Significantly stronger electric fields can be found in heavy atoms and molecules\,\cite{Hinds97a}, in which case the hidden momentum might be significantly larger.

In the calculation leading to (\ref{finalpmulti}), the emergence of the hidden momentum can be traced to the `$1$' in the spin-orbit factor $f_k$ (a translating magnetic-dipole moment resembles an electric-dipole moment\,\cite{Uhlenbeck26a, Barone73a, Muller92a}) whilst the emergence of the momentum difference $m_k(\hat{\bar{\mathbf{v}}}_k-\hat{\mathbf{v}}_k)$ can be traced to the `$-q_k/2 m_k \gamma_k$' (Thomas precession\,\cite{Thomas26a, Thomas27a, Gunther-Mohr54a, Barone73a, Muller92a}). This seems natural, as the position difference $\hat{\pmb{\delta}}_k$ is intimately associated with Thomas precession\,\cite{Barone73a,Muller92a}.

For a more detailed discussion of the energy, linear momentum, angular momentum and boost momentum of a molecule to order $1/c^2$ see\,\cite{Cameron18a}.

\subsection{Total hidden momentum and its cancellation}
We recognise $\hat{\mathbf{P}}=\sum_k \hat{\mathbf{p}}_k$ \label{Ptotal} as being the total momentum of the molecule. $\hat{\mathbf{P}}$ is conserved and generates (Cartesian\,\cite{Cameron15a}) translations of the molecule in space\,\cite{Noether18a, Bessel-Hagen21a}. The hidden contribution to $\hat{\mathbf{P}}$ is countered by an equal and opposite contribution due to the magnetic-dipole moments of the particles,
\begin{equation}
-\sum_k \frac{\hat{\mathbf{m}}_k\times \pmb{\nabla}_k\hat{\Phi}_k^q}{c^2}+\sum_k q_k \hat{\mathbf{A}}^\mathbf{m}_k=0.
\end{equation}
Thus, the total hidden momentum of the molecule is cancelled by the momentum of the intramolecular electromagnetic field, as one might expect\,\cite{Thomson04a,Thomson04b,Thomson04c, Griffiths99, Babson09a}.


\section{Outlook}
\label{Outlook}
An electron vortex\,\cite{Bliokh11a,Larocque16a,Bialynicki-Birula17a,Barnett17a,Bliokh17aa,Kruining17a,Lloyd17a,Bliokh17a} in an electric field $\mathbf{E}$ might harbour a hidden momentum due to a modification of the electron's orbital motion by $\mathbf{E}$, in addition to the spin-based hidden momentum identified in this paper. The orbital-based hidden momentum should take the form $\mathbf{m}_\ell\times\mathbf{E}/c^2$, with $\mathbf{m}_\ell$ the orbital magnetic-dipole moment of the electron. Assuming that $|\mathbf{m}_\ell|\sim -q\hbar|\ell|/2m$, this is $\lesssim -q\hbar|\ell||\mathbf{E}|/2mc^2=|\ell||\mathbf{E}|\times 10^{-40}\textrm{A}.\textrm{s}^2$ in magnitude. The orbital-based hidden momentum could be significantly larger than the spin-based hidden momentum (expectation value $\lesssim -q\hbar|\mathbf{E}|/2mc^2$ in magnitude), as the orbital angular momentum quantum number $\ell\in\{0,\pm1,\dots\}$ is unbounded. 

Inferring the existence of hidden momentum in the laboratory is an interesting problem. One might endeavour to measure the associated \textit{angular} momentum, which is not necessarily cancelled by the angular momentum of the field that gives rise to the hidden momentum - unlike the total \textit{linear} momentum, the total \textit{angular} momentum of a system `at rest' need not vanish \cite{Griffiths99}. An electron vortex with a large orbital angular momentum, perturbed by an electric field, might prove particularly suitable for this purpose.

The hidden momentum of a system like the one described in the introduction might be referred to more descriptively as a hidden \textit{kinetic} momentum, to emphasise that it is an imbalance of the \textit{kinetic} momenta of the system's constituent particles: `$\sum\gamma m \mathbf{v}\ne0$'\,\cite{Shockley67a, Griffiths99, Babson09a, Filho15a}. In this paper we have established that even a single particle like the electron can harbour a hidden momentum associated with its spin. We can now conceive, therefore, of systems containing such particles in which there is \textit{no} imbalance of the \textit{kinetic} momenta of the particles and yet the total \textit{hidden} momentum of the particles is non-zero: `$\sum\gamma m \mathbf{v}=0$' but `$\sum\mathbf{m}\times\mathbf{E}/c^2 \ne 0$'. One might say that such a system harbours hidden \textit{hidden} momentum, in distinction to hidden \textit{kinetic} momentum. A loop of electric current (driven through a resistive element by a battery) encircling the tip of a (long) magnetised needle is one such system. To appreciate this, consider a simple model of such a system in which the loop is circular and lies in the $x-y$ plane whilst the tip of the needle coincides with the centre of the loop, at the origin. If we imagine that the magnetic-dipole moment `$\mathbf{m}$' of each charge carrier is aligned radially due to the magnetic field of the needle whilst the electric field `$\mathbf{E}$' driving the current around the loop is aligned azimuthally, then the hidden momentum `$\mathbf{m}\times\mathbf{E}/c^2$' of each charge carrier is aligned axially. Thus, the system harbours a hidden \textit{hidden} momentum `$\sum\mathbf{m}\times\mathbf{E}/c^2 \ne 0$', with no hidden \textit{kinetic} momentum to speak of: `$\sum\gamma m \mathbf{v}=0$'. Hall effects\,\cite{Hall79a,Dyakonov71a} have been neglected in our argument. We do not expect these to dramatically alter the underlying physics, however.

\section{Acknowledgements}
This work was supported by the EPSRC (EP/M004694/1) and The Leverhulme Trust (RPG-2017-048). We thank Gergely Ferenczi for his advice and encouragement.


\begin{appendix}
\section{Representations}
\label{Representations}
The Foldy-Wouthuysen representation was introduced in\,\cite{Foldy50a} by Foldy and Wouthuysen to establish a correspondence between Dirac's fully relativistic theory expressed in the Dirac representation\,\cite{Dirac28a} and the low-energy Pauli description of spin $1/2$ particles, familiar from atomic and molecular studies for example\,\cite{Pauli27a} - it is not obvious that the low-energy limit of the former coincides with the latter. In this paper we use both the Dirac and Foldy-Wouthuysen representations because some quantities such as the position $\hat{\mathbf{r}}_q'=\mathbf{r}$ of charge have simple operator representatives in the Dirac representation whilst others such as the mean position $\hat{\mathbf{r}}_M=\mathbf{r}$ have simple operator representatives in the Foldy-Wouthuysen representation instead. The following is a summary of key results from\,\cite{Foldy50a}. 

For a free electron, the Foldy-Wouthuysen representation is related to the Dirac representation by the unitary operator
\begin{eqnarray}
U&=&\textrm{exp}\left\{\textrm{i}\left[-\frac{\textrm{i}\beta\pmb{\alpha}\cdot\mathbf{p}}{2p}  \textrm{tan}^{-1}\left(\frac{p}{mc}\right)\right]\right\}.
\end{eqnarray}
The transformed Hamiltonian
\begin{equation}
\hat{H}=U\hat{H}'U^\dagger=\beta E_p
\end{equation}
is diagonal and even: the upper and lower components of the transformed spinor $\psi=U_0\psi'$ correspond, respectively, to positive and negative energies. 

For an electron in an external electromagnetic field, the Foldy-Wouthuysen representation is instead related to the Dirac representation by a sequence of unitary transformations. Taking
\begin{equation}
U^\dagger=\textrm{e}^{-\textrm{i}S_1}\textrm{e}^{-\textrm{i}S_2}\textrm{e}^{-\textrm{i}S_3}\dots, 
\end{equation}
with
\begin{eqnarray}
S_1&=& -\frac{\textrm{i} \beta\pmb{\alpha}\cdot(\mathbf{p}-q\mathbf{A})}{2mc}, \\ 
S_2 &=& \frac{\hbar q \pmb{\alpha}\cdot\mathbf{E}}{4 m^2c^3},\\
S_3 &=& \frac{\textrm{i}\beta \alpha_a\alpha_b\alpha_c(p_a-qA_a)(p_b-qA_b)(p_c-qA_c)}{6m^3c^3}, \nonumber \\
&& 
\end{eqnarray}
gives
\begin{eqnarray}
\hat{H}&=&U\hat{H}'U^\dagger - \textrm{i}\hbar U\frac{\partial U^\dagger }{\partial t} \nonumber \\
&=&\beta m c^2+\frac{\beta p^2}{2 m} +q\Phi -\frac{\beta p^4}{8m^3c^2}+\frac{\hbar^2 q \nabla^2\Phi}{8m^2 c^2} \nonumber \\
&-&\frac{q \mathbf{s}\cdot (\mathbf{p}\times\pmb{\nabla}\Phi)}{2m^2c^2}-\frac{\beta q \mathbf{p}\cdot\mathbf{A}}{m}-\frac{q \beta(\mathbf{s}\cdot\mathbf{B})}{m}+\textrm{O}(\tfrac{1}{c^3}) \nonumber \\
&&
\end{eqnarray}
as the transformed Hamiltonian, which is even to order $1/c^2$. 

\end{appendix}


\bibliographystyle{apsrev4-1}
\bibliography{HiddenMomentum}

\begin{thebibliography}{69}%
\makeatletter
\providecommand \@ifxundefined [1]{%
 \@ifx{#1\undefined}
}%
\providecommand \@ifnum [1]{%
 \ifnum #1\expandafter \@firstoftwo
 \else \expandafter \@secondoftwo
 \fi
}%
\providecommand \@ifx [1]{%
 \ifx #1\expandafter \@firstoftwo
 \else \expandafter \@secondoftwo
 \fi
}%
\providecommand \natexlab [1]{#1}%
\providecommand \enquote  [1]{``#1''}%
\providecommand \bibnamefont  [1]{#1}%
\providecommand \bibfnamefont [1]{#1}%
\providecommand \citenamefont [1]{#1}%
\providecommand \href@noop [0]{\@secondoftwo}%
\providecommand \href [0]{\begingroup \@sanitize@url \@href}%
\providecommand \@href[1]{\@@startlink{#1}\@@href}%
\providecommand \@@href[1]{\endgroup#1\@@endlink}%
\providecommand \@sanitize@url [0]{\catcode `\\12\catcode `\$12\catcode
  `\&12\catcode `\#12\catcode `\^12\catcode `\_12\catcode `\%12\relax}%
\providecommand \@@startlink[1]{}%
\providecommand \@@endlink[0]{}%
\providecommand \url  [0]{\begingroup\@sanitize@url \@url }%
\providecommand \@url [1]{\endgroup\@href {#1}{\urlprefix }}%
\providecommand \urlprefix  [0]{URL }%
\providecommand \Eprint [0]{\href }%
\providecommand \doibase [0]{http://dx.doi.org/}%
\providecommand \selectlanguage [0]{\@gobble}%
\providecommand \bibinfo  [0]{\@secondoftwo}%
\providecommand \bibfield  [0]{\@secondoftwo}%
\providecommand \translation [1]{[#1]}%
\providecommand \BibitemOpen [0]{}%
\providecommand \bibitemStop [0]{}%
\providecommand \bibitemNoStop [0]{.\EOS\space}%
\providecommand \EOS [0]{\spacefactor3000\relax}%
\providecommand \BibitemShut  [1]{\csname bibitem#1\endcsname}%
\let\auto@bib@innerbib\@empty
\bibitem [{\citenamefont {Shockley}\ and\ \citenamefont
  {James}(1967)}]{Shockley67a}%
  \BibitemOpen
  \bibfield  {author} {\bibinfo {author} {\bibfnamefont {W.}~\bibnamefont
  {Shockley}}\ and\ \bibinfo {author} {\bibfnamefont {R.~P.}\ \bibnamefont
  {James}},\ }\href@noop {} {\bibfield  {journal} {\bibinfo  {journal} {Phys.
  Rev. Lett.}\ }\textbf {\bibinfo {volume} {18}},\ \bibinfo {pages} {876}
  (\bibinfo {year} {1967})}\BibitemShut {NoStop}%
\bibitem [{\citenamefont {Griffiths}(1999)}]{Griffiths99}%
  \BibitemOpen
  \bibfield  {author} {\bibinfo {author} {\bibfnamefont {D.~J.}\ \bibnamefont
  {Griffiths}},\ }\href@noop {} {\emph {\bibinfo {title} {Introduction to
  Electrodynamics}}}\ (\bibinfo  {publisher} {Pearson Education},\ \bibinfo
  {year} {1999})\BibitemShut {NoStop}%
\bibitem [{\citenamefont {Babson}\ \emph {et~al.}(2009)\citenamefont {Babson},
  \citenamefont {Reynolds}, \citenamefont {Bjorkquist},\ and\ \citenamefont
  {Griffiths}}]{Babson09a}%
  \BibitemOpen
  \bibfield  {author} {\bibinfo {author} {\bibfnamefont {D.}~\bibnamefont
  {Babson}}, \bibinfo {author} {\bibfnamefont {S.~P.}\ \bibnamefont
  {Reynolds}}, \bibinfo {author} {\bibfnamefont {R.}~\bibnamefont
  {Bjorkquist}}, \ and\ \bibinfo {author} {\bibfnamefont {D.~J.}\ \bibnamefont
  {Griffiths}},\ }\href@noop {} {\bibfield  {journal} {\bibinfo  {journal} {Am.
  J. Phys.}\ }\textbf {\bibinfo {volume} {77}},\ \bibinfo {pages} {826}
  (\bibinfo {year} {2009})}\BibitemShut {NoStop}%
\bibitem [{\citenamefont {Filho}\ and\ \citenamefont
  {Saldanha}(2015)}]{Filho15a}%
  \BibitemOpen
  \bibfield  {author} {\bibinfo {author} {\bibfnamefont {J.~S.~O.}\
  \bibnamefont {Filho}}\ and\ \bibinfo {author} {\bibfnamefont {P.~L.}\
  \bibnamefont {Saldanha}},\ }\href@noop {} {\bibfield  {journal} {\bibinfo
  {journal} {Phys. Rev. A}\ }\textbf {\bibinfo {volume} {92}},\ \bibinfo
  {pages} {052107} (\bibinfo {year} {2015})}\BibitemShut {NoStop}%
\bibitem [{\citenamefont {Thomson}(1904{\natexlab{a}})}]{Thomson04a}%
  \BibitemOpen
  \bibfield  {author} {\bibinfo {author} {\bibfnamefont {J.~J.}\ \bibnamefont
  {Thomson}},\ }\href@noop {} {\emph {\bibinfo {title} {Electricity and
  Matter}}}\ (\bibinfo  {publisher} {Charles Scribner's Sons},\ \bibinfo {year}
  {1904})\BibitemShut {NoStop}%
\bibitem [{\citenamefont {Thomson}(1904{\natexlab{b}})}]{Thomson04b}%
  \BibitemOpen
  \bibfield  {author} {\bibinfo {author} {\bibfnamefont {J.~J.}\ \bibnamefont
  {Thomson}},\ }\href@noop {} {\bibfield  {journal} {\bibinfo  {journal} {Phil.
  Mag.}\ }\textbf {\bibinfo {volume} {8}},\ \bibinfo {pages} {331} (\bibinfo
  {year} {1904}{\natexlab{b}})}\BibitemShut {NoStop}%
\bibitem [{\citenamefont {Thomson}(1904{\natexlab{c}})}]{Thomson04c}%
  \BibitemOpen
  \bibfield  {author} {\bibinfo {author} {\bibfnamefont {J.~J.}\ \bibnamefont
  {Thomson}},\ }\href@noop {} {\emph {\bibinfo {title} {Elements of the
  Mathematical Theory of Electricity and Magnetism}}}\ (\bibinfo  {publisher}
  {Cambridge University Press},\ \bibinfo {year} {1904})\BibitemShut {NoStop}%
\bibitem [{\citenamefont {van Vleck}\ and\ \citenamefont
  {Huang}(1969)}]{vanVleck69a}%
  \BibitemOpen
  \bibfield  {author} {\bibinfo {author} {\bibfnamefont {J.~H.}\ \bibnamefont
  {van Vleck}}\ and\ \bibinfo {author} {\bibfnamefont {N.~L.}\ \bibnamefont
  {Huang}},\ }\href@noop {} {\bibfield  {journal} {\bibinfo  {journal} {Phys.
  Lett.}\ }\textbf {\bibinfo {volume} {28A}},\ \bibinfo {pages} {768} (\bibinfo
  {year} {1969})}\BibitemShut {NoStop}%
\bibitem [{\citenamefont {Bliokh}\ \emph {et~al.}(2011)\citenamefont {Bliokh},
  \citenamefont {Dennis},\ and\ \citenamefont {Nori}}]{Bliokh11a}%
  \BibitemOpen
  \bibfield  {author} {\bibinfo {author} {\bibfnamefont {K.~Y.}\ \bibnamefont
  {Bliokh}}, \bibinfo {author} {\bibfnamefont {M.~R.}\ \bibnamefont {Dennis}},
  \ and\ \bibinfo {author} {\bibfnamefont {F.}~\bibnamefont {Nori}},\
  }\href@noop {} {\bibfield  {journal} {\bibinfo  {journal} {Phys. Rev. Lett.}\
  }\textbf {\bibinfo {volume} {107}},\ \bibinfo {pages} {174802} (\bibinfo
  {year} {2011})}\BibitemShut {NoStop}%
\bibitem [{\citenamefont {Larocque}\ \emph {et~al.}(2016)\citenamefont
  {Larocque}, \citenamefont {Bouchard}, \citenamefont {Grillo}, \citenamefont
  {Sit}, \citenamefont {Frabboni}, \citenamefont {Dunin-Borkowski},
  \citenamefont {Padgett}, \citenamefont {Boyd},\ and\ \citenamefont
  {Karimi}}]{Larocque16a}%
  \BibitemOpen
  \bibfield  {author} {\bibinfo {author} {\bibfnamefont {H.}~\bibnamefont
  {Larocque}}, \bibinfo {author} {\bibfnamefont {F.}~\bibnamefont {Bouchard}},
  \bibinfo {author} {\bibfnamefont {V.}~\bibnamefont {Grillo}}, \bibinfo
  {author} {\bibfnamefont {A.}~\bibnamefont {Sit}}, \bibinfo {author}
  {\bibfnamefont {S.}~\bibnamefont {Frabboni}}, \bibinfo {author}
  {\bibfnamefont {R.~E.}\ \bibnamefont {Dunin-Borkowski}}, \bibinfo {author}
  {\bibfnamefont {M.~J.}\ \bibnamefont {Padgett}}, \bibinfo {author}
  {\bibfnamefont {R.~W.}\ \bibnamefont {Boyd}}, \ and\ \bibinfo {author}
  {\bibfnamefont {E.}~\bibnamefont {Karimi}},\ }\href@noop {} {\bibfield
  {journal} {\bibinfo  {journal} {Phys. Rev. Lett.}\ }\textbf {\bibinfo
  {volume} {117}},\ \bibinfo {pages} {154801} (\bibinfo {year}
  {2016})}\BibitemShut {NoStop}%
\bibitem [{\citenamefont {Bialynicki-Birula}\ and\ \citenamefont
  {Bialynicki-Birula}(2017)}]{Bialynicki-Birula17a}%
  \BibitemOpen
  \bibfield  {author} {\bibinfo {author} {\bibfnamefont {I.}~\bibnamefont
  {Bialynicki-Birula}}\ and\ \bibinfo {author} {\bibfnamefont {Z.}~\bibnamefont
  {Bialynicki-Birula}},\ }\href@noop {} {\bibfield  {journal} {\bibinfo
  {journal} {Phys. Rev. Lett.}\ }\textbf {\bibinfo {volume} {118}},\ \bibinfo
  {pages} {114801} (\bibinfo {year} {2017})}\BibitemShut {NoStop}%
\bibitem [{\citenamefont {Barnett}(2017)}]{Barnett17a}%
  \BibitemOpen
  \bibfield  {author} {\bibinfo {author} {\bibfnamefont {S.~M.}\ \bibnamefont
  {Barnett}},\ }\href@noop {} {\bibfield  {journal} {\bibinfo  {journal} {Phys.
  Rev. Lett.}\ }\textbf {\bibinfo {volume} {118}},\ \bibinfo {pages} {114802}
  (\bibinfo {year} {2017})}\BibitemShut {NoStop}%
\bibitem [{\citenamefont {Bliokh}\ \emph
  {et~al.}(2017{\natexlab{a}})\citenamefont {Bliokh}, \citenamefont {Ivanov},
  \citenamefont {Guzzinati}, \citenamefont {Clark}, \citenamefont {van Boxem},
  \citenamefont {B\'{e}ch\'{e}}, \citenamefont {Juchtmans}, \citenamefont
  {Alonso}, \citenamefont {Schattschneider}, \citenamefont {Nori},\ and\
  \citenamefont {Verbeeck}}]{Bliokh17aa}%
  \BibitemOpen
  \bibfield  {author} {\bibinfo {author} {\bibfnamefont {K.~Y.}\ \bibnamefont
  {Bliokh}}, \bibinfo {author} {\bibfnamefont {I.~P.}\ \bibnamefont {Ivanov}},
  \bibinfo {author} {\bibfnamefont {G.}~\bibnamefont {Guzzinati}}, \bibinfo
  {author} {\bibfnamefont {L.}~\bibnamefont {Clark}}, \bibinfo {author}
  {\bibfnamefont {R.}~\bibnamefont {van Boxem}}, \bibinfo {author}
  {\bibfnamefont {A.}~\bibnamefont {B\'{e}ch\'{e}}}, \bibinfo {author}
  {\bibfnamefont {R.}~\bibnamefont {Juchtmans}}, \bibinfo {author}
  {\bibfnamefont {M.~A.}\ \bibnamefont {Alonso}}, \bibinfo {author}
  {\bibfnamefont {P.}~\bibnamefont {Schattschneider}}, \bibinfo {author}
  {\bibfnamefont {F.}~\bibnamefont {Nori}}, \ and\ \bibinfo {author}
  {\bibfnamefont {J.}~\bibnamefont {Verbeeck}},\ }\href@noop {} {\bibfield
  {journal} {\bibinfo  {journal} {Phys. Rep.}\ }\textbf {\bibinfo {volume}
  {690}},\ \bibinfo {pages} {1} (\bibinfo {year}
  {2017}{\natexlab{a}})}\BibitemShut {NoStop}%
\bibitem [{\citenamefont {van Kruining}\ \emph {et~al.}(2017)\citenamefont {van
  Kruining}, \citenamefont {Hayrapetyan},\ and\ \citenamefont
  {G\"{o}tte}}]{Kruining17a}%
  \BibitemOpen
  \bibfield  {author} {\bibinfo {author} {\bibfnamefont {K.}~\bibnamefont {van
  Kruining}}, \bibinfo {author} {\bibfnamefont {A.~G.}\ \bibnamefont
  {Hayrapetyan}}, \ and\ \bibinfo {author} {\bibfnamefont {J.~B.}\ \bibnamefont
  {G\"{o}tte}},\ }\href@noop {} {\bibfield  {journal} {\bibinfo  {journal}
  {Phys. Rev. Lett.}\ }\textbf {\bibinfo {volume} {119}},\ \bibinfo {pages}
  {030401} (\bibinfo {year} {2017})}\BibitemShut {NoStop}%
\bibitem [{\citenamefont {Lloyd}\ \emph {et~al.}(2017)\citenamefont {Lloyd},
  \citenamefont {Babiker}, \citenamefont {Thirunavukkarasu},\ and\
  \citenamefont {Yuan}}]{Lloyd17a}%
  \BibitemOpen
  \bibfield  {author} {\bibinfo {author} {\bibfnamefont {S.~M.}\ \bibnamefont
  {Lloyd}}, \bibinfo {author} {\bibfnamefont {M.}~\bibnamefont {Babiker}},
  \bibinfo {author} {\bibfnamefont {G.}~\bibnamefont {Thirunavukkarasu}}, \
  and\ \bibinfo {author} {\bibfnamefont {J.}~\bibnamefont {Yuan}},\ }\href@noop
  {} {\bibfield  {journal} {\bibinfo  {journal} {Rev. Mod. Phys.}\ }\textbf
  {\bibinfo {volume} {89}},\ \bibinfo {pages} {035004} (\bibinfo {year}
  {2017})}\BibitemShut {NoStop}%
\bibitem [{\citenamefont {Bliokh}\ \emph
  {et~al.}(2017{\natexlab{b}})\citenamefont {Bliokh}, \citenamefont {Dennis},\
  and\ \citenamefont {Nori}}]{Bliokh17a}%
  \BibitemOpen
  \bibfield  {author} {\bibinfo {author} {\bibfnamefont {K.~Y.}\ \bibnamefont
  {Bliokh}}, \bibinfo {author} {\bibfnamefont {M.~R.}\ \bibnamefont {Dennis}},
  \ and\ \bibinfo {author} {\bibfnamefont {F.}~\bibnamefont {Nori}},\
  }\href@noop {} {\bibfield  {journal} {\bibinfo  {journal} {Phys. Rev. A}\
  }\textbf {\bibinfo {volume} {96}},\ \bibinfo {pages} {023622} (\bibinfo
  {year} {2017}{\natexlab{b}})}\BibitemShut {NoStop}%
\bibitem [{\citenamefont {Dirac}(1928)}]{Dirac28a}%
  \BibitemOpen
  \bibfield  {author} {\bibinfo {author} {\bibfnamefont {P.~A.~M.}\
  \bibnamefont {Dirac}},\ }\href@noop {} {\bibfield  {journal} {\bibinfo
  {journal} {Proc. Roy. Soc. A}\ }\textbf {\bibinfo {volume} {117}},\ \bibinfo
  {pages} {610} (\bibinfo {year} {1928})}\BibitemShut {NoStop}%
\bibitem [{\citenamefont {Foldy}\ and\ \citenamefont
  {Wouthuysen}(1950)}]{Foldy50a}%
  \BibitemOpen
  \bibfield  {author} {\bibinfo {author} {\bibfnamefont {L.~L.}\ \bibnamefont
  {Foldy}}\ and\ \bibinfo {author} {\bibfnamefont {S.~A.}\ \bibnamefont
  {Wouthuysen}},\ }\href@noop {} {\bibfield  {journal} {\bibinfo  {journal}
  {Phys. Rev.}\ }\textbf {\bibinfo {volume} {78}},\ \bibinfo {pages} {29}
  (\bibinfo {year} {1950})}\BibitemShut {NoStop}%
\bibitem [{\citenamefont {Born}\ and\ \citenamefont {Infeld}(1935)}]{Born35a}%
  \BibitemOpen
  \bibfield  {author} {\bibinfo {author} {\bibfnamefont {M.}~\bibnamefont
  {Born}}\ and\ \bibinfo {author} {\bibfnamefont {L.}~\bibnamefont {Infeld}},\
  }\href@noop {} {\bibfield  {journal} {\bibinfo  {journal} {Proc. Roy. Soc.
  A}\ }\textbf {\bibinfo {volume} {150}},\ \bibinfo {pages} {141} (\bibinfo
  {year} {1935})}\BibitemShut {NoStop}%
\bibitem [{\citenamefont {Pryce}(1935)}]{Pryce35a}%
  \BibitemOpen
  \bibfield  {author} {\bibinfo {author} {\bibfnamefont {M.~H.~L.}\
  \bibnamefont {Pryce}},\ }\href@noop {} {\bibfield  {journal} {\bibinfo
  {journal} {Proc. Roy. Soc. A}\ }\textbf {\bibinfo {volume} {150}},\ \bibinfo
  {pages} {166} (\bibinfo {year} {1935})}\BibitemShut {NoStop}%
\bibitem [{\citenamefont {Pryce}(1948)}]{Pryce48a}%
  \BibitemOpen
  \bibfield  {author} {\bibinfo {author} {\bibfnamefont {M.~H.~L.}\
  \bibnamefont {Pryce}},\ }\href@noop {} {\bibfield  {journal} {\bibinfo
  {journal} {Proc. Roy. Soc. A}\ }\textbf {\bibinfo {volume} {195}},\ \bibinfo
  {pages} {62} (\bibinfo {year} {1948})}\BibitemShut {NoStop}%
\bibitem [{\citenamefont {Barut}\ and\ \citenamefont {Malin}(1968)}]{Barut68a}%
  \BibitemOpen
  \bibfield  {author} {\bibinfo {author} {\bibfnamefont {A.~O.}\ \bibnamefont
  {Barut}}\ and\ \bibinfo {author} {\bibfnamefont {S.}~\bibnamefont {Malin}},\
  }\href@noop {} {\bibfield  {journal} {\bibinfo  {journal} {Rev. Mod. Phys.}\
  }\textbf {\bibinfo {volume} {40}},\ \bibinfo {pages} {632} (\bibinfo {year}
  {1968})}\BibitemShut {NoStop}%
\bibitem [{\citenamefont {Barone}(1973)}]{Barone73a}%
  \BibitemOpen
  \bibfield  {author} {\bibinfo {author} {\bibfnamefont {S.~R.}\ \bibnamefont
  {Barone}},\ }\href@noop {} {\bibfield  {journal} {\bibinfo  {journal} {Phys.
  Rev. D}\ }\textbf {\bibinfo {volume} {8}},\ \bibinfo {pages} {3492} (\bibinfo
  {year} {1973})}\BibitemShut {NoStop}%
\bibitem [{\citenamefont {Barut}\ and\ \citenamefont
  {Bracken}(1981)}]{Barut81a}%
  \BibitemOpen
  \bibfield  {author} {\bibinfo {author} {\bibfnamefont {A.~O.}\ \bibnamefont
  {Barut}}\ and\ \bibinfo {author} {\bibfnamefont {A.~J.}\ \bibnamefont
  {Bracken}},\ }\href@noop {} {\bibfield  {journal} {\bibinfo  {journal} {Phys.
  Rev. D}\ }\textbf {\bibinfo {volume} {23}},\ \bibinfo {pages} {2454}
  (\bibinfo {year} {1981})}\BibitemShut {NoStop}%
\bibitem [{\citenamefont {Bliokh}\ and\ \citenamefont
  {Nori}(2012)}]{Bliokh12a}%
  \BibitemOpen
  \bibfield  {author} {\bibinfo {author} {\bibfnamefont {K.~Y.}\ \bibnamefont
  {Bliokh}}\ and\ \bibinfo {author} {\bibfnamefont {F.}~\bibnamefont {Nori}},\
  }\href@noop {} {\bibfield  {journal} {\bibinfo  {journal} {Phys. Rev. Lett.}\
  }\textbf {\bibinfo {volume} {108}},\ \bibinfo {pages} {120403} (\bibinfo
  {year} {2012})}\BibitemShut {NoStop}%
\bibitem [{\citenamefont {Smirnova}\ \emph {et~al.}(2018)\citenamefont
  {Smirnova}, \citenamefont {Tavin}, \citenamefont {Bliokh},\ and\
  \citenamefont {Nori}}]{Smirnova18a}%
  \BibitemOpen
  \bibfield  {author} {\bibinfo {author} {\bibfnamefont {D.~A.}\ \bibnamefont
  {Smirnova}}, \bibinfo {author} {\bibfnamefont {V.~M.}\ \bibnamefont {Tavin}},
  \bibinfo {author} {\bibfnamefont {K.~Y.}\ \bibnamefont {Bliokh}}, \ and\
  \bibinfo {author} {\bibfnamefont {F.}~\bibnamefont {Nori}},\ }\href@noop {}
  {\bibfield  {journal} {\bibinfo  {journal} {arXiv:1711.03255v2}\ } (\bibinfo
  {year} {2018})}\BibitemShut {NoStop}%
\bibitem [{\citenamefont {Schr\"{o}dinger}(1930)}]{Schrodinger30a}%
  \BibitemOpen
  \bibfield  {author} {\bibinfo {author} {\bibfnamefont {E.}~\bibnamefont
  {Schr\"{o}dinger}},\ }\href@noop {} {\bibfield  {journal} {\bibinfo
  {journal} {Sitzungsberichte der Preussischen Akademie der Wissenschaften
  Physikalisch-Mathematische Klasse}\ }\textbf {\bibinfo {volume} {24}},\
  \bibinfo {pages} {418} (\bibinfo {year} {1930})}\BibitemShut {NoStop}%
\bibitem [{\citenamefont {Schr\"{o}dinger}(1931)}]{Schrodinger31a}%
  \BibitemOpen
  \bibfield  {author} {\bibinfo {author} {\bibfnamefont {E.}~\bibnamefont
  {Schr\"{o}dinger}},\ }\href@noop {} {\bibfield  {journal} {\bibinfo
  {journal} {Sitzungsberichte der Preussischen Akademie der Wissenschaften
  Physikalisch-Mathematische Klasse}\ }\textbf {\bibinfo {volume} {3}},\
  \bibinfo {pages} {63} (\bibinfo {year} {1931})}\BibitemShut {NoStop}%
\bibitem [{\citenamefont {Giambiagi}(1960)}]{Giambiagi60a}%
  \BibitemOpen
  \bibfield  {author} {\bibinfo {author} {\bibfnamefont {J.~J.}\ \bibnamefont
  {Giambiagi}},\ }\href@noop {} {\bibfield  {journal} {\bibinfo  {journal} {Il
  Nuovo Cimento}\ }\textbf {\bibinfo {volume} {16}},\ \bibinfo {pages} {202}
  (\bibinfo {year} {1960})}\BibitemShut {NoStop}%
\bibitem [{\citenamefont {Saavedra}(1965)}]{Saavedra65a}%
  \BibitemOpen
  \bibfield  {author} {\bibinfo {author} {\bibfnamefont {I.}~\bibnamefont
  {Saavedra}},\ }\href@noop {} {\bibfield  {journal} {\bibinfo  {journal}
  {Nucl. Phys.}\ }\textbf {\bibinfo {volume} {74}},\ \bibinfo {pages} {677}
  (\bibinfo {year} {1965})}\BibitemShut {NoStop}%
\bibitem [{\citenamefont {Pauli}(1927)}]{Pauli27a}%
  \BibitemOpen
  \bibfield  {author} {\bibinfo {author} {\bibfnamefont {W.}~\bibnamefont
  {Pauli}},\ }\href@noop {} {\bibfield  {journal} {\bibinfo  {journal}
  {Zeitschrift f\"{u}r Physik}\ }\textbf {\bibinfo {volume} {37}},\ \bibinfo
  {pages} {601} (\bibinfo {year} {1927})}\BibitemShut {NoStop}%
\bibitem [{\citenamefont {Eddington}(1929)}]{Eddington29a}%
  \BibitemOpen
  \bibfield  {author} {\bibinfo {author} {\bibfnamefont {A.~S.}\ \bibnamefont
  {Eddington}},\ }\href@noop {} {\bibfield  {journal} {\bibinfo  {journal}
  {Proc. Roy. Soc. A}\ }\textbf {\bibinfo {volume} {122}},\ \bibinfo {pages}
  {358} (\bibinfo {year} {1929})}\BibitemShut {NoStop}%
\bibitem [{\citenamefont {Breit}(1929)}]{Breit29a}%
  \BibitemOpen
  \bibfield  {author} {\bibinfo {author} {\bibfnamefont {G.}~\bibnamefont
  {Breit}},\ }\href@noop {} {\bibfield  {journal} {\bibinfo  {journal} {Phys.
  Rev.}\ }\textbf {\bibinfo {volume} {34}},\ \bibinfo {pages} {553} (\bibinfo
  {year} {1929})}\BibitemShut {NoStop}%
\bibitem [{\citenamefont {Huang}(1952)}]{Huang52a}%
  \BibitemOpen
  \bibfield  {author} {\bibinfo {author} {\bibfnamefont {K.}~\bibnamefont
  {Huang}},\ }\href@noop {} {\bibfield  {journal} {\bibinfo  {journal} {Am. J.
  Phys.}\ }\textbf {\bibinfo {volume} {20}},\ \bibinfo {pages} {479} (\bibinfo
  {year} {1952})}\BibitemShut {NoStop}%
\bibitem [{\citenamefont {Muller}(1992)}]{Muller92a}%
  \BibitemOpen
  \bibfield  {author} {\bibinfo {author} {\bibfnamefont {R.~A.}\ \bibnamefont
  {Muller}},\ }\href@noop {} {\bibfield  {journal} {\bibinfo  {journal} {Am. J.
  Phys.}\ }\textbf {\bibinfo {volume} {60}},\ \bibinfo {pages} {313} (\bibinfo
  {year} {1992})}\BibitemShut {NoStop}%
\bibitem [{\citenamefont {Darwin}(1932)}]{Darwin32a}%
  \BibitemOpen
  \bibfield  {author} {\bibinfo {author} {\bibfnamefont {C.~G.}\ \bibnamefont
  {Darwin}},\ }\href@noop {} {\bibfield  {journal} {\bibinfo  {journal} {Proc.
  R. Soc. Lond. A}\ }\textbf {\bibinfo {volume} {136}},\ \bibinfo {pages} {36}
  (\bibinfo {year} {1932})}\BibitemShut {NoStop}%
\bibitem [{\citenamefont {van Enk}\ and\ \citenamefont
  {Nienhuis}(1994{\natexlab{a}})}]{vanEnk94a}%
  \BibitemOpen
  \bibfield  {author} {\bibinfo {author} {\bibfnamefont {S.~J.}\ \bibnamefont
  {van Enk}}\ and\ \bibinfo {author} {\bibfnamefont {G.}~\bibnamefont
  {Nienhuis}},\ }\href@noop {} {\bibfield  {journal} {\bibinfo  {journal}
  {Europhys. Lett.}\ }\textbf {\bibinfo {volume} {25}},\ \bibinfo {pages} {497}
  (\bibinfo {year} {1994}{\natexlab{a}})}\BibitemShut {NoStop}%
\bibitem [{\citenamefont {van Enk}\ and\ \citenamefont
  {Nienhuis}(1994{\natexlab{b}})}]{vanEnk94b}%
  \BibitemOpen
  \bibfield  {author} {\bibinfo {author} {\bibfnamefont {S.~J.}\ \bibnamefont
  {van Enk}}\ and\ \bibinfo {author} {\bibfnamefont {G.}~\bibnamefont
  {Nienhuis}},\ }\href@noop {} {\bibfield  {journal} {\bibinfo  {journal} {J.
  Mod. Opt.}\ }\textbf {\bibinfo {volume} {41}},\ \bibinfo {pages} {963}
  (\bibinfo {year} {1994}{\natexlab{b}})}\BibitemShut {NoStop}%
\bibitem [{\citenamefont {Barnett}(2010)}]{Barnett10a}%
  \BibitemOpen
  \bibfield  {author} {\bibinfo {author} {\bibfnamefont {S.~M.}\ \bibnamefont
  {Barnett}},\ }\href@noop {} {\bibfield  {journal} {\bibinfo  {journal} {J.
  Mod. Opt.}\ }\textbf {\bibinfo {volume} {57}},\ \bibinfo {pages} {1339}
  (\bibinfo {year} {2010})}\BibitemShut {NoStop}%
\bibitem [{\citenamefont {Bliokh}\ \emph {et~al.}(2010)\citenamefont {Bliokh},
  \citenamefont {Alonso}, \citenamefont {Ostrovskaya},\ and\ \citenamefont
  {Aiello}}]{Bliokh10a}%
  \BibitemOpen
  \bibfield  {author} {\bibinfo {author} {\bibfnamefont {K.~Y.}\ \bibnamefont
  {Bliokh}}, \bibinfo {author} {\bibfnamefont {M.~A.}\ \bibnamefont {Alonso}},
  \bibinfo {author} {\bibfnamefont {E.~A.}\ \bibnamefont {Ostrovskaya}}, \ and\
  \bibinfo {author} {\bibfnamefont {A.}~\bibnamefont {Aiello}},\ }\href@noop {}
  {\bibfield  {journal} {\bibinfo  {journal} {Phys. Rev. A}\ }\textbf {\bibinfo
  {volume} {82}},\ \bibinfo {pages} {063825} (\bibinfo {year}
  {2010})}\BibitemShut {NoStop}%
\bibitem [{\citenamefont {Bialynicki-Birula}\ and\ \citenamefont
  {Bialynicka-Birula}(2011)}]{Ivo11a}%
  \BibitemOpen
  \bibfield  {author} {\bibinfo {author} {\bibfnamefont {I.}~\bibnamefont
  {Bialynicki-Birula}}\ and\ \bibinfo {author} {\bibfnamefont {Z.}~\bibnamefont
  {Bialynicka-Birula}},\ }\href@noop {} {\bibfield  {journal} {\bibinfo
  {journal} {J. Opt.}\ }\textbf {\bibinfo {volume} {13}},\ \bibinfo {pages}
  {064014} (\bibinfo {year} {2011})}\BibitemShut {NoStop}%
\bibitem [{\citenamefont {Adams}\ and\ \citenamefont
  {Blount}(1959)}]{Adams59a}%
  \BibitemOpen
  \bibfield  {author} {\bibinfo {author} {\bibfnamefont {E.~N.}\ \bibnamefont
  {Adams}}\ and\ \bibinfo {author} {\bibfnamefont {E.~I.}\ \bibnamefont
  {Blount}},\ }\href@noop {} {\bibfield  {journal} {\bibinfo  {journal} {J.
  Phys. Chem. Solids}\ }\textbf {\bibinfo {volume} {10}},\ \bibinfo {pages}
  {286} (\bibinfo {year} {1959})}\BibitemShut {NoStop}%
\bibitem [{\citenamefont {Bliokh}(2005)}]{Bliokh05a}%
  \BibitemOpen
  \bibfield  {author} {\bibinfo {author} {\bibfnamefont {K.~Y.}\ \bibnamefont
  {Bliokh}},\ }\href@noop {} {\bibfield  {journal} {\bibinfo  {journal}
  {Europhys. Lett.}\ }\textbf {\bibinfo {volume} {72}},\ \bibinfo {pages} {7}
  (\bibinfo {year} {2005})}\BibitemShut {NoStop}%
\bibitem [{\citenamefont {B\'{e}rard}\ and\ \citenamefont
  {Mohrbach}(2006)}]{Berard06a}%
  \BibitemOpen
  \bibfield  {author} {\bibinfo {author} {\bibfnamefont {A.}~\bibnamefont
  {B\'{e}rard}}\ and\ \bibinfo {author} {\bibfnamefont {H.}~\bibnamefont
  {Mohrbach}},\ }\href@noop {} {\bibfield  {journal} {\bibinfo  {journal}
  {Phys. Lett. A}\ }\textbf {\bibinfo {volume} {352}},\ \bibinfo {pages} {190}
  (\bibinfo {year} {2006})}\BibitemShut {NoStop}%
\bibitem [{\citenamefont {Chang}\ and\ \citenamefont {Niu}(2008)}]{Chang08a}%
  \BibitemOpen
  \bibfield  {author} {\bibinfo {author} {\bibfnamefont {M.~C.}\ \bibnamefont
  {Chang}}\ and\ \bibinfo {author} {\bibfnamefont {Q.}~\bibnamefont {Niu}},\
  }\href@noop {} {\bibfield  {journal} {\bibinfo  {journal} {J. Phys.: Condens.
  Matter}\ }\textbf {\bibinfo {volume} {20}},\ \bibinfo {pages} {193202}
  (\bibinfo {year} {2008})}\BibitemShut {NoStop}%
\bibitem [{\citenamefont {Xiao}\ \emph {et~al.}(2010)\citenamefont {Xiao},
  \citenamefont {Chang},\ and\ \citenamefont {Niu}}]{Xiao10a}%
  \BibitemOpen
  \bibfield  {author} {\bibinfo {author} {\bibfnamefont {D.}~\bibnamefont
  {Xiao}}, \bibinfo {author} {\bibfnamefont {M.~C.}\ \bibnamefont {Chang}}, \
  and\ \bibinfo {author} {\bibfnamefont {Q.}~\bibnamefont {Niu}},\ }\href@noop
  {} {\bibfield  {journal} {\bibinfo  {journal} {Rev. Mod. Phys.}\ }\textbf
  {\bibinfo {volume} {82}},\ \bibinfo {pages} {1959} (\bibinfo {year}
  {2010})}\BibitemShut {NoStop}%
\bibitem [{\citenamefont {Takahashi}\ and\ \citenamefont
  {Nagaosa}(2015)}]{Takahashi15a}%
  \BibitemOpen
  \bibfield  {author} {\bibinfo {author} {\bibfnamefont {R.}~\bibnamefont
  {Takahashi}}\ and\ \bibinfo {author} {\bibfnamefont {N.}~\bibnamefont
  {Nagaosa}},\ }\href@noop {} {\bibfield  {journal} {\bibinfo  {journal} {Phys.
  Rev. B}\ }\textbf {\bibinfo {volume} {91}},\ \bibinfo {pages} {245133}
  (\bibinfo {year} {2015})}\BibitemShut {NoStop}%
\bibitem [{\citenamefont {Maxwell}(1861)}]{Maxwell61a}%
  \BibitemOpen
  \bibfield  {author} {\bibinfo {author} {\bibfnamefont {J.~C.}\ \bibnamefont
  {Maxwell}},\ }\href@noop {} {\bibfield  {journal} {\bibinfo  {journal} {Phil.
  Mag.}\ }\textbf {\bibinfo {volume} {21}},\ \bibinfo {pages} {281 } (\bibinfo
  {year} {1861})}\BibitemShut {NoStop}%
\bibitem [{\citenamefont {Lorentz}(1895)}]{Lorentz95a}%
  \BibitemOpen
  \bibfield  {author} {\bibinfo {author} {\bibfnamefont {H.~A.}\ \bibnamefont
  {Lorentz}},\ }\href@noop {} {\emph {\bibinfo {title} {Versuch einer Theorie
  der electrischen und optischen und optischen Erscheinungen in bewegten
  K\"{o}rpern}}}\ (\bibinfo  {publisher} {E. J. Brill},\ \bibinfo {year}
  {1895})\BibitemShut {NoStop}%
\bibitem [{\citenamefont {Darwin}(1928)}]{Darwin28a}%
  \BibitemOpen
  \bibfield  {author} {\bibinfo {author} {\bibfnamefont {C.~G.}\ \bibnamefont
  {Darwin}},\ }\href@noop {} {\bibfield  {journal} {\bibinfo  {journal} {Proc.
  Roy. Soc. A}\ }\textbf {\bibinfo {volume} {118}},\ \bibinfo {pages} {654}
  (\bibinfo {year} {1928})}\BibitemShut {NoStop}%
\bibitem [{\citenamefont {Itoh}(1963)}]{Itoh65a}%
  \BibitemOpen
  \bibfield  {author} {\bibinfo {author} {\bibfnamefont {T.}~\bibnamefont
  {Itoh}},\ }\href@noop {} {\bibfield  {journal} {\bibinfo  {journal} {Rev.
  Mod. Phys.}\ }\textbf {\bibinfo {volume} {37}},\ \bibinfo {pages} {159}
  (\bibinfo {year} {1963})}\BibitemShut {NoStop}%
\bibitem [{\citenamefont {Pachucki}\ and\ \citenamefont
  {Karshenboim}(1995)}]{Pachucki95a}%
  \BibitemOpen
  \bibfield  {author} {\bibinfo {author} {\bibfnamefont {K.}~\bibnamefont
  {Pachucki}}\ and\ \bibinfo {author} {\bibfnamefont {S.~G.}\ \bibnamefont
  {Karshenboim}},\ }\href@noop {} {\bibfield  {journal} {\bibinfo  {journal}
  {J. Phys. B: At. Mol. Opt. Phys.}\ }\textbf {\bibinfo {volume} {28}},\
  \bibinfo {pages} {L221} (\bibinfo {year} {1995})}\BibitemShut {NoStop}%
\bibitem [{\citenamefont {Helgaker}\ \emph {et~al.}()\citenamefont {Helgaker},
  \citenamefont {J{\o}rgensen},\ and\ \citenamefont {Olsen}}]{Helgaker}%
  \BibitemOpen
  \bibfield  {author} {\bibinfo {author} {\bibfnamefont {T.}~\bibnamefont
  {Helgaker}}, \bibinfo {author} {\bibfnamefont {P.}~\bibnamefont
  {J{\o}rgensen}}, \ and\ \bibinfo {author} {\bibfnamefont {J.}~\bibnamefont
  {Olsen}},\ }\href@noop {} {\bibinfo  {journal}
  {{http://folk.uio.no/helgaker/talks/Hamiltonian.pdf}}\ }\BibitemShut
  {NoStop}%
\bibitem [{\citenamefont {Uhlenbeck}\ and\ \citenamefont
  {Goudsmit}(1926)}]{Uhlenbeck26a}%
  \BibitemOpen
\bibfield  {journal} {  }\bibfield  {author} {\bibinfo {author} {\bibfnamefont
  {G.~E.}\ \bibnamefont {Uhlenbeck}}\ and\ \bibinfo {author} {\bibfnamefont
  {S.}~\bibnamefont {Goudsmit}},\ }\href@noop {} {\bibfield  {journal}
  {\bibinfo  {journal} {Nature}\ }\textbf {\bibinfo {volume} {117}},\ \bibinfo
  {pages} {264} (\bibinfo {year} {1926})}\BibitemShut {NoStop}%
\bibitem [{\citenamefont {Thomas}(1926)}]{Thomas26a}%
  \BibitemOpen
  \bibfield  {author} {\bibinfo {author} {\bibfnamefont {L.~H.}\ \bibnamefont
  {Thomas}},\ }\href@noop {} {\bibfield  {journal} {\bibinfo  {journal}
  {Nature}\ }\textbf {\bibinfo {volume} {117}},\ \bibinfo {pages} {514}
  (\bibinfo {year} {1926})}\BibitemShut {NoStop}%
\bibitem [{\citenamefont {Thomas}(1927)}]{Thomas27a}%
  \BibitemOpen
  \bibfield  {author} {\bibinfo {author} {\bibfnamefont {L.~H.}\ \bibnamefont
  {Thomas}},\ }\href@noop {} {\bibfield  {journal} {\bibinfo  {journal} {Phil.
  Mag.}\ }\textbf {\bibinfo {volume} {3}},\ \bibinfo {pages} {1} (\bibinfo
  {year} {1927})}\BibitemShut {NoStop}%
\bibitem [{\citenamefont {Gunther-Mohr}\ \emph {et~al.}(1954)\citenamefont
  {Gunther-Mohr}, \citenamefont {Townes},\ and\ \citenamefont {van
  Vleck}}]{Gunther-Mohr54a}%
  \BibitemOpen
  \bibfield  {author} {\bibinfo {author} {\bibfnamefont {G.~R.}\ \bibnamefont
  {Gunther-Mohr}}, \bibinfo {author} {\bibfnamefont {C.~H.}\ \bibnamefont
  {Townes}}, \ and\ \bibinfo {author} {\bibfnamefont {J.~H.}\ \bibnamefont {van
  Vleck}},\ }\href@noop {} {\bibfield  {journal} {\bibinfo  {journal} {Phys.
  Rev.}\ }\textbf {\bibinfo {volume} {94}},\ \bibinfo {pages} {1191} (\bibinfo
  {year} {1954})}\BibitemShut {NoStop}%
\bibitem [{\citenamefont {Breit}(1930)}]{Breit30a}%
  \BibitemOpen
  \bibfield  {author} {\bibinfo {author} {\bibfnamefont {G.}~\bibnamefont
  {Breit}},\ }\href@noop {} {\bibfield  {journal} {\bibinfo  {journal} {Phys.
  Rev.}\ }\textbf {\bibinfo {volume} {36}},\ \bibinfo {pages} {383} (\bibinfo
  {year} {1930})}\BibitemShut {NoStop}%
\bibitem [{\citenamefont {Breit}(1932)}]{Breit32a}%
  \BibitemOpen
  \bibfield  {author} {\bibinfo {author} {\bibfnamefont {G.}~\bibnamefont
  {Breit}},\ }\href@noop {} {\bibfield  {journal} {\bibinfo  {journal} {Phys.
  Rev.}\ }\textbf {\bibinfo {volume} {39}},\ \bibinfo {pages} {616} (\bibinfo
  {year} {1932})}\BibitemShut {NoStop}%
\bibitem [{\citenamefont {Chraplyvy}(1953{\natexlab{a}})}]{Chraplyvy53a}%
  \BibitemOpen
  \bibfield  {author} {\bibinfo {author} {\bibfnamefont {Z.~V.}\ \bibnamefont
  {Chraplyvy}},\ }\href@noop {} {\bibfield  {journal} {\bibinfo  {journal}
  {Phys. Rev.}\ }\textbf {\bibinfo {volume} {91}},\ \bibinfo {pages} {388}
  (\bibinfo {year} {1953}{\natexlab{a}})}\BibitemShut {NoStop}%
\bibitem [{\citenamefont {Chraplyvy}(1953{\natexlab{b}})}]{Chraplyvy53b}%
  \BibitemOpen
  \bibfield  {author} {\bibinfo {author} {\bibfnamefont {Z.~V.}\ \bibnamefont
  {Chraplyvy}},\ }\href@noop {} {\bibfield  {journal} {\bibinfo  {journal}
  {Phys. Rev.}\ }\textbf {\bibinfo {volume} {92}},\ \bibinfo {pages} {1310}
  (\bibinfo {year} {1953}{\natexlab{b}})}\BibitemShut {NoStop}%
\bibitem [{Note1()}]{Note1}%
  \BibitemOpen
  \bibinfo {note} {For the $j$th nucleus, the kinetic position $\protect
  \mathaccentV {hat}05E{\protect \mathaccentV {bar}016{\protect \mathbf
  {r}}}_j$ appears to differ from the centre $\protect \mathaccentV
  {hat}05E{\protect \mathbf {r}}_j+(2m_j\gamma _j/q_j-1)\protect \mathaccentV
  {hat}05E{\protect \pmb {\delta }}_j$ of charge.}\BibitemShut {Stop}%
\bibitem [{\citenamefont {Hinds}(1997)}]{Hinds97a}%
  \BibitemOpen
  \bibfield  {author} {\bibinfo {author} {\bibfnamefont {E.~A.}\ \bibnamefont
  {Hinds}},\ }\href@noop {} {\bibfield  {journal} {\bibinfo  {journal} {Phys.
  Scripta.}\ }\textbf {\bibinfo {volume} {T70}},\ \bibinfo {pages} {34}
  (\bibinfo {year} {1997})}\BibitemShut {NoStop}%
\bibitem [{\citenamefont {Cameron}\ and\ \citenamefont
  {Cotter}(2018)}]{Cameron18a}%
  \BibitemOpen
  \bibfield  {author} {\bibinfo {author} {\bibfnamefont {R.~P.}\ \bibnamefont
  {Cameron}}\ and\ \bibinfo {author} {\bibfnamefont {J.~P.}\ \bibnamefont
  {Cotter}},\ }\href@noop {} {\bibfield  {journal} {\bibinfo  {journal} {J.
  Phys. B: At. Mol. Opt. Phys.}\ }\textbf {\bibinfo {volume} {51}},\ \bibinfo
  {pages} {105101} (\bibinfo {year} {2018})}\BibitemShut {NoStop}%
\bibitem [{\citenamefont {Cameron}\ \emph {et~al.}(2015)\citenamefont
  {Cameron}, \citenamefont {Speirits}, \citenamefont {Gilson}, \citenamefont
  {Allen},\ and\ \citenamefont {Barnett}}]{Cameron15a}%
  \BibitemOpen
  \bibfield  {author} {\bibinfo {author} {\bibfnamefont {R.~P.}\ \bibnamefont
  {Cameron}}, \bibinfo {author} {\bibfnamefont {F.~C.}\ \bibnamefont
  {Speirits}}, \bibinfo {author} {\bibfnamefont {C.~R.}\ \bibnamefont
  {Gilson}}, \bibinfo {author} {\bibfnamefont {L.}~\bibnamefont {Allen}}, \
  and\ \bibinfo {author} {\bibfnamefont {S.~M.}\ \bibnamefont {Barnett}},\
  }\href@noop {} {\bibfield  {journal} {\bibinfo  {journal} {J. Opt.}\ }\textbf
  {\bibinfo {volume} {17}},\ \bibinfo {pages} {125610} (\bibinfo {year}
  {2015})}\BibitemShut {NoStop}%
\bibitem [{\citenamefont {Noether}(1918)}]{Noether18a}%
  \BibitemOpen
  \bibfield  {author} {\bibinfo {author} {\bibfnamefont {E.}~\bibnamefont
  {Noether}},\ }\href@noop {} {\bibfield  {journal} {\bibinfo  {journal}
  {Nachrichten von der Gesellschaft der Wissenschaften zu G\"{o}ttingen,
  Mathematisch-Physikalische Klasse}\ }\textbf {\bibinfo {volume} {2}},\
  \bibinfo {pages} {235} (\bibinfo {year} {1918})}\BibitemShut {NoStop}%
\bibitem [{\citenamefont {Bessel-Hagen}(1921)}]{Bessel-Hagen21a}%
  \BibitemOpen
  \bibfield  {author} {\bibinfo {author} {\bibfnamefont {E.}~\bibnamefont
  {Bessel-Hagen}},\ }\href@noop {} {\bibfield  {journal} {\bibinfo  {journal}
  {Mathematische Annalen}\ }\textbf {\bibinfo {volume} {84}},\ \bibinfo {pages}
  {258} (\bibinfo {year} {1921})}\BibitemShut {NoStop}%
\bibitem [{\citenamefont {Hall}(1879)}]{Hall79a}%
  \BibitemOpen
  \bibfield  {author} {\bibinfo {author} {\bibfnamefont {E.~H.}\ \bibnamefont
  {Hall}},\ }\href@noop {} {\bibfield  {journal} {\bibinfo  {journal} {Am. J.
  Math.}\ }\textbf {\bibinfo {volume} {2}},\ \bibinfo {pages} {287} (\bibinfo
  {year} {1879})}\BibitemShut {NoStop}%
\bibitem [{\citenamefont {Dyakonov}\ and\ \citenamefont
  {Perel'}(1971)}]{Dyakonov71a}%
  \BibitemOpen
  \bibfield  {author} {\bibinfo {author} {\bibfnamefont {M.~I.}\ \bibnamefont
  {Dyakonov}}\ and\ \bibinfo {author} {\bibfnamefont {V.~I.}\ \bibnamefont
  {Perel'}},\ }\href@noop {} {\bibfield  {journal} {\bibinfo  {journal} {Sov.
  Phys. JETP Lett.}\ }\textbf {\bibinfo {volume} {13}},\ \bibinfo {pages} {467}
  (\bibinfo {year} {1971})}\BibitemShut {NoStop}%
\end{thebibliography}%


\end{document}